\documentstyle[12pt]{article}
\def\beq{\begin{equation}}
\def\eeq{\end{equation}}

\begin{document}

\begin{titlepage}
\begin{center}
{\Large \bf Theoretical Physics Institute \\
University of Minnesota \\}  \end{center}
\vspace{0.3in}
\begin{flushright}
TPI-MINN-92/56-T \\
October 1992
\end{flushright}
\vspace{0.4in}
\begin{center}
{\Large \bf Zeros of tree-level amplitudes at multi-boson thresholds \\}
\vspace{0.2in}
{\bf M.B. Voloshin  \\ }
Theoretical Physics Institute, University of Minnesota \\
Minneapolis, MN 55455 \\
and \\
Institute of Theoretical and Experimental Physics  \\
                         Moscow, 117259 \\
\vspace{0.2in}
{\bf   Abstract  \\ }
\end{center}
Propagation of particles with emission of arbitrary number of identical
bosons all being at rest is considered. It is shown that
in certain models the tree-level amplitudes for production of $n$ scalar
bosons by two incoming particles are all equal to zero at the threshold
starting from some small number $n$. In particular this nullification occurs
for production of massive scalars by two Goldstone bosons in the linear
sigma model for $n > 1$ and also for production of Higgs bosons in the
Standard Model by gauge bosons and/or by fermions, provided that the ratia
of their masses to that of the Higgs boson take special discrete values.

\end{titlepage}

Based on the technique recently suggested by Lowell
Brown$^{\cite{brown}}$ it has been found$^{\cite{v,smith}}$ that the
on-mass-shell scattering amplitudes in the $\lambda \phi^4$ theory
for production of $n$ particles of the field $\phi$ by two incoming
ones display a peculiar behavior at the $n$
particle threshold. Namely, the tree-level result for those amplitudes
is strictly zero for $n > 4$ in the theory without spontaneous symmetry
breaking (positive mass term)$^{\cite{v}}$ and for $n >2$ in the theory
with spontaneous symmetry breaking (negative mass
term)$^{\cite{smith}}$.

In this paper few more examples of the same
behavior are given, where the initial particles are different from the
bosons being produced. The processes considered can be generically
written as $\chi \,\chi \to n\, \phi$ and $f\, {\bar f} \to n\, \phi$ where
$\chi$ stands for either a different than $\phi$ scalar boson or a massive
gauge boson and $f$ stands for a fermion.  In particular it is found
that in a linear sigma model the amplitude of production of $n$
massive $\sigma$ particles by two incoming Goldstone bosons vanishes at
the threshold for any $n$ larger than $1$. As a consequence of this
nullification the amplitude of production of more than one Higgs boson
by longitudinal massive vectors in the Standard Model is vanishing at
the threshold to the leading order in the ratio of the vector boson and
the Higgs boson masses. This behavior for the particular case $n=2$ was
observed in some previous direct calculations$^{\cite{zero2}}$.
Furthermore for the transversal vector bosons it will be demonstrated
that if the ratio of the masses satisfies the relation

\beq
4 m_V^2/m_H^2 = N(N+1)
\label{vmagic}
\eeq
with $N$ being an integer, the tree-level amplitudes of the processes
$V_T\,V_T \to n\, H$ are all zero at the threshold for $n > N$. For the
fermions in the Standard Model the ``magic" mass ratio is of the form

\beq
m_f/m_H = N/2
\label{fmagic}
\eeq
with an integer $N$, in which case the nullification at the threshold
occurs for the amplitudes of the processes $f {\bar f} \to n\, H$ for
all $n$ greater than $N$ and also for $n=N$. Notice that both in the case of
transversal vector bosons and in the case of fermions this implies that the
nullification of the production amplitudes occurs in fact at all
kinematically possible thresholds for on-mass-shell processes due to the
obvious constraint $2m_V \le n\,m_H$ or $2m_f \le n\,m_H$, provided of
course that the relation (\ref{vmagic}) or (\ref{fmagic}) holds.

To arrive at these conclusions we calculate at the tree level the propagator
$D_n(p)$ of a quantum particle with emission of $n$ on-mass-shell bosons of
the field $\phi$ all being at rest and $p$ being the final momentum in the
propagator after the emission (see Fig. 1). As a most direct approach to
this calculation we use a combination of the Brown's technique and the one
based on the generating function method$^{\cite{akp}}$ of solving recursion
relations$^{\cite{v1,v2}}$ for the propagators $D(p)$. To formulate the
recursion relations we introduce the notation $-i V(\phi)$ for the vertex of
interaction of the propagating particle ($\chi$ or $f$) with arbitrary
external field $\phi$ and for the beginning we consider the propagator
of a bosonic field $\chi$ whose mass in the absence of interaction with
$\phi$ is $m_{0\chi}$. The recursion equations for the propagators
$D_n(p)$ arise from graph shown in Fig. 2 and can be written as

\beq
D_n(p)=-i\,d(p+nq) \sum_{n_1} {{n!} \over {n_1!\,(n-n_1)!}} V_{n_1}
D_{n-n_1}(p) ~,
\label{recur}
\eeq
where $q$ is the momentum of each of the produced $\phi$-bosons, in their
rest frame $q= (M, {\bf 0})$ with $M$ being the mass of the $\phi$ boson,
$d(p)$ is the propagator in the absence of interaction with the field
$\phi$: $ d(p)= i / (p^2-m_{0\chi}^2)$,  and $V_{n}$
are the matrix elements for production of $n$ bosons at rest by the operator
$V(\phi(x))$:

\beq
V_n= \langle n | V(\phi(0)) | 0 \rangle ~.
\label{mate}
\eeq
Notice, that in a theory where the field $\phi$ develops a vacuum
expectation value $v= \langle 0 | \phi | 0 \rangle$ the ``bare"
propagator $d(p)$ does not coincide with the propagator

\beq
D_0(p)= {i \over {p^2-m_{0\chi}^2-V_0}}
\label{d0cor}
\eeq
since the physical mass of the $\chi$ boson receives also a contribution
from $V_0=V(v)$: $m_\chi^2=m_{0\chi}^2+V_0$.

The recursion equations (\ref{recur}) are converted into a differential
equation by introducing generating functions ${\cal V}(z)$ and ${\cal
D}(p\,;~z)$ related to the $V_n$ and $D_n(p)$ as

\beq
{\cal V}(z)=\sum_{n=0} {{z^n} \over {n!}} V_n ~~~~~{\rm and}~~~~~
{\cal D}(p\,;~z)=-i\,\sum_{n=0} {{z^n} \over {n!}} D_n(p)~.
\label{gf}
\eeq
In terms of the generating functions the equation (\ref{recur}) is
equivalent to the $n$-th term of the Taylor series expansion of the
differential equation

\beq
\left ( M^2 z^2 {{d^2} \over {d z^2}} + (2 (p \cdot q)+M^2 ) z {{d} \over
{dz}} + p^2 - m_{0\chi}^2 - {\cal V}(z) \right ) {\cal D}(p\,;~ z) =1~
\label{difur}
\eeq
with the initial condition at $z=0$ that ${\cal D}(p\,;~0)=D_0(p)$
(the consistency of this condition is guaranteed by the inhomogeneuos term
in the right hand side) and that the solution has Taylor expansion in
positive powers of $z$. Alternatively one can specify the latter condition
by calculating explicitly the propagator $D_1(p)$, i.e. with emission of one
particle, which determines the first derivative $d{\cal D}(p\,;~z)/dz$ at
$z=0$.

At the tree level the generating function ${\cal V}(z)$ is determined by the
generating function $\Phi (z)$ for the matrix elements of the field $\phi$
itself,
\beq
\Phi(z)= \sum \langle n |\phi(0) |0 \rangle z^n/n!~,
\label{gphi}
\eeq
${\cal V}(z) = V(\Phi(z))$. The function $\Phi(z)$ is known for a class of
theories$^{\cite{v1,akp,brown}}$. In this paper the analysis is restricted
to a sub-class of probably the most physical interest i.e. where the
Lagrangian density for the field $\phi$ is that of the $\lambda \phi^4$
theory

\beq
{\cal L}={1 \over 2} (\partial_\mu \phi)^2 -{1 \over 2} m^2 \phi^2 - {1
\over 4} \lambda \phi^4
\label{lagr}
\eeq
with or without spontaneous symmetry breaking, i.e. with either sign of
the $m^2$. For the case of no symmetry breaking one
has$^{\cite{v1,akp,brown}}$

\beq
\Phi(z)={z \over {1- (\lambda /8M^2) z^2}}
\label{nossb}
\eeq
with $M=m$ and for the case of negative $m^2$ the generating function is
given by$^{\cite{akp,brown}}$

\beq
\Phi(z)= v {{1+z/2v} \over {1-z/2v}} ~,
\label{ssb}
\eeq
where $v=|m|/\sqrt{\lambda}$ and also the physical mass of the bosons of the
field $\phi$ is $M=\sqrt{2}\,|m|$.

We shall further restrict the generality of the present discussion by
requiring that the vertex function $V(\phi)$ has the simple form

\beq
V(\phi)= \xi \phi^2
\label{inter}
\eeq
which still covers quite a few interesting cases.

Let us first consider the unbroken symmetry case. Upon substitution
into equation (\ref{difur}) of the generating function ${\cal V}(z)$ as
determined by eqs.(\ref{inter}) and (\ref{nossb}) the equation takes the
form

\beq
\left ( M^2 z^2 {{d^2} \over {d z^2}} + (2 (p \cdot q)+M^2 ) z {{d} \over
{dz}} + p^2 - m_{0\chi}^2 - {{\xi z^2} \over {[1-(\lambda/8M^2)\,z^2]^2}}
\right ) {\cal D}(p\,;~ z) =1~.
\label{d1nossb}
\eeq
It is convenient to introduce the notation $\epsilon=(p \cdot
q)/M^2$ and $\omega=\sqrt{\epsilon^2-(p^2-m_{0 \chi}^2)/M^2}$, where the
square root of a positive number is taken to be positive.
In the rest frame of the produced bosons $\epsilon$ is the energy
$p_0$ of the final particle in the propagator in units of $M$ and $\omega^2$
is related to its spatial momentum ${\bf p}$, namely it is ${\bf p}^2+
m_{0\chi}^2$ in units of $M^2$.  Thus the difference $\epsilon^2-\omega^2$
is the measure of by how much the momentum $p$ is off shell.

The solution of the equation (\ref{d1nossb}) can be sought in the form

\beq
{\cal D}(p\,;~z)= y^{-\epsilon/2} \, f(p\,;~y)~,
\label{solform}
\eeq
where $y = -\lambda z^2/(8
M^2)$, in terms of which eq.(\ref{d1nossb}) is rewritten as

\beq
\left ( 4 y^2 {{d^2} \over {dy^2}} + 4 y { d \over {dy}} - \omega^2  +
8 {\xi \over \lambda} {y \over {(1+y)^2}} \right ) f(p\,;~y) = {{y^{\epsilon
/2}} \over {M^2}}~.
\label{d2nossb}
\eeq

For the broken symmetry case, using in the same manner the generating
function (\ref{ssb}) instead of (\ref{nossb}) and seeking the solution
in the form (\ref{solform}) in terms of the variable $y=-z/(2v)$, one
arrives at the equation essentially identical to eq.(\ref{d2nossb}) up to
rescaling of some terms:

\beq
\left ( 4 y^2 {{d^2} \over {dy^2}} + 4 y { d \over {dy}} - 4{\bar \omega}^2
+ 8 {\xi \over \lambda} {y \over {(1+y)^2}} \right ) f(p\,;~y) = 4
{{y^{\epsilon /2}} \over {M^2}}~.
\label{d2ssb}
\eeq
where ${\bar \omega}^2= \epsilon^2 -(p^2-m_{0\chi}^2-\xi v^2)/M^2$, and it
can be also reminded that the expression for $M$ in terms of $|m|$ differs
from that in the case of unbroken symmetry by factor $\sqrt{2}$. Therefore
the solution of the equation (\ref{d2ssb}) is obtained from that of
eq.(\ref{d2nossb}) by replacing $\omega \to 2 {\bar \omega}$ and by overall
rescaling of the solution by the factor 4.

The operator in the homogeneous left hand side of eq.(\ref{d2nossb}) is
related to the well known exactly solvable Schr\"odinger operator with the
P\"oschl-Teller potential.  Namely, the substitution $y=e^{2 \tau}$
converts the operator into

\beq
{{d^2} \over {d \tau^2}} -\omega^2 + { {s(s+1)} \over {(\cosh{\tau})^2}}
\label{oper}
\eeq
with $s(s+1)=2 \xi/\lambda$.

The properties of the operator (\ref{oper}) are well known from Quantum
Mechanics (see e.g.  in the textbook \cite{ll}). In particular these
properties are quite special when the parameter $s$ is integer, $s=N$, and
there is no reflection of waves in the one-dimensional Quantum Mechanical
problem. It is in this special case of integer $s$ that the infinite series
of zeros of the threshold amplitudes arises and which will be considered
here.  The solution of the equation (\ref{d1nossb}) for ${\cal D}(p\,;~z)$ can
be written explicitly$^{\cite{ll}}$ in terms of the hypergeometric function
$F(-s, \omega-s,\omega+1,-y)$, which in the case $s=N$ is reduced to a
Jacobi polynomial of the $N$-th power. We therefore define the function

\beq
F_N(\omega,y) = {{\Gamma(1+\omega)} \over {\Gamma(N+1+\omega)}}
y^{-\omega/2} (1+y)^{N+1} {{d^N} \over {dy^N}} \left ( {{y^{\omega+N}}\over
{(1+y)^{N+1}}} \right )
\label{myfunc}
\eeq
and write the explicit formula for ${\cal D}(p\,;~z)$ in the unbroken symmetry
case in terms of the variable $y$ in the form

\beq
{\cal D}(p, z(y))= {{y^{-\epsilon/2}} \over {4 \omega M^2}} \left [
F_N(-\omega, y) \int_0^y u^{{\epsilon \over 2} - 1} F_N(\omega,u)\, du  +
F_N(\omega, y) \int_y^\infty u^{{\epsilon \over 2} - 1} F_N(-\omega,u)\, du
\right ]~,
\label{solut}
\eeq
where $y$ is assumed to be positive. Notice, that $F_N(\omega, y)$ is
regular at small $y$: $F_N(\omega, y) \approx y^{\omega/2}$, while
$F_N(-\omega,y)$ is singular at $y \to 0$. At $y \to \infty$ the behavior of
these functions is switched. Also in its dependence on $\omega$ the function
$F_N(-\omega,y)$ develops simple poles at $\omega = 1,~2, \ldots ,~N$ due to
the factor $\Gamma(1-\omega) /\Gamma(N+1- \omega)$ in its definition.

We can now proceed to considering the amplitude of production $n$
$\phi$-bosons by two incoming particles. In terms of the propagator $D_n(p)$
this amplitude corresponds to negative $p_0$ (the final line in the
propagator is that of an incoming particle), so that $\epsilon=-n/2$. The
on-mass-shell amplitude is given by the double pole of this propagator when
$\omega=-\epsilon=n/2$. The only possibility for the expression in the right
hand side of the equation (\ref{solut}) to develop a double pole for
positive $\omega$ is when one pole term comes from the function
$F_N(-\omega,y)$ and the other one comes from the divergence of the integral.
At negative $\epsilon$ the first of the integrals in eq.(\ref{solut}) indeed
develops single poles, so the double poles are possible only for the values
$\omega = 1,~2, \ldots ,~N~$, i.e. only as long as $n \le 2N$ and thus the
on-mass-shell threshold amplitudes are all zero for $n > 2N$

For the case of broken symmetry as we have seen the generating function
${\cal D}$ for the propagators is determined by the same operator as for the
unbroken symmetry case with $\omega$ rescaled by factor 2
(eq.(\ref{d2ssb})). Therefore the same consideration leads one to the
conclusion that in this case the on-mass-shell threshold amplitudes are zero
for all $n$ larger than $N$.

This nullification of the amplitudes at the thresholds is the one observed
for production of the $\phi$ bosons by particles of the same field, whose
propagation is described by the interaction (\ref{inter}) with $\xi=3
\lambda$, so that $2 \xi/\lambda=6$ and thus $N=2$ both in the case of no
spontaneous symmetry breaking$^{\cite{v,v2}}$ and in the case of broken
symmetry$^{\cite{smith}}$. Here we consider few more applications
corresponding to different integer values $s=N$. \\[0.2in]

\newpage
{\bf Linear sigma model.}

In the Lagrangian density of the linear sigma model

\beq
{\cal L}={1 \over 2} (\partial_\mu \sigma)^2 + {1 \over 2} (\partial_\mu
{\vec \pi})^2 - {\lambda \over 4} (\sigma^2+{\vec \pi}^2 - v^2)^2
\label{lsigma}
\eeq
the field $\sigma$ plays the role of $\phi$ and the $\pi$ components play
the role of $\chi$ in the previous discussion.  The interaction between the
``pions" and the $\sigma$ in terms of our previous definition (\ref{inter})
corresponds to $\xi=\lambda$, thus $2 \xi/\lambda=2$, so that $N=1$.
Therefore one immediately concludes that all the tree-level amplitudes of
the processes $2 \pi \to n \, \sigma$ vanish at the threshold except for
$n=1$.\\[0.2in]

{\bf Higgs boson production by massive vector bosons.}

For a heavy Higgs boson in the standard model to the leading order in the
ratio of the masses $m_V/m_H$ the amplitudes of production of Higgs bosons
by longitudinally polarized massive vector bosons are given by those of the
linear sigma model, where the $\pi$ components describe the
longitudinally polarized gauge bosons. Therefore from the previous paragraph
one concludes that in this order the threshold amplitudes for $V_L\,V_L \to
n \, H$ vanish at the threshold for $n > 1$. For the particular case $n=2$
this behavior was observed in direct calculations$^{\cite{zero2}}$ of
Feynman graphs.

For the transversal components of the gauge bosons the Lagrangian in the
Proca gauge is equivalent to that of a scalar interacting with the properly
normalized Higgs field $\phi$ with the constant such that $2 \xi /\lambda =
4\,m_V^2/m_H^2$. Therefore we conclude that if this mass ratio has a value
corresponding to the equation (\ref{vmagic}) the tree-level amplitudes for
the processes $V_T\,V_T \to n\, H$ vanish at the threshold for $n > N$.
Notice however that for the on-mass-shell processes of this type only the
values of $n$ larger than $N$ are possible, due to the kinematical
constraint $2 m_V \le n\, m_H$. Therefore for all physically possible
on-mass-shell processes of this kind the amplitudes vanish at the
threshold.\\[0.2in]

{\bf Production of scalar bosons by fermions.}

Let us turn now to the case when the $\phi$ bosons are produced by fermions
due to Yukawa interaction corresponding to the vertex function

\beq
V(\phi)=h \phi
\label{finter}
\eeq
where $h$ is the coupling constant\footnote{It is assumed for simplicity
that the coupling is scalar, though the following discussion can be readily
generalized to the case of arbitrary mixture of scalar and pseudoscalar
couplings.}. The analog of the bosonic equation (\ref{difur}) for the
fermion propagator generating function reads as

\beq
\left ( (\gamma \cdot q) z { d \over {dz}} + (\gamma \cdot p) - m_{0 f} - h
\Phi(z) \right ) {\cal D}(p\,;~z)=1~,
\label{fdifur}
\eeq
where $m_{0 f}$ is the fermion mass term in the absence of interaction with
$\phi$, and naturally the generating function ${\cal D}$ in this case is a
matrix acting on bispinors.

The solution of the equation (\ref{fdifur}) can be sought in the form

\beq
{\cal D}(p\,;~z)=\left ( (\gamma \cdot q) z { d \over {dz}} + (\gamma \cdot p)
+ m_{0 f} + h \Phi(z) \right ) {\cal F}(p\,;~z)~,
\label{fform}
\eeq
and the equation for the function ${\cal F}(p\,;~z)$ thus reads as

\beq
\left ( M^2 z^2 {{d^2} \over {d z^2}} + (2 (p \cdot q)+M^2 ) z {{d} \over
{dz}} + p^2 -(m_{0f}+ h \Phi(z))^2 + (\gamma \cdot q) h z {{d \Phi(z)}
\over { dz}} \right ) {\cal F}(p\,;~z)=1~.
\label{fd1}
\eeq
This equation
takes a familiar from the previous discussion form in the case relevant to
the Standard Model, when $m_{0f}=0$, i.e. when the fermion gets all of its
mass due to the spontaneous symmetry breaking. Substituting then the
explicit expression (\ref{ssb}) for $\Phi(z)$ one finds that the potential
term in the equation (\ref{fd1}) has the form

\beq
-h^2 \Phi(z)^2 + (\gamma \cdot q) h z (d \Phi(z)/dz) = -m_f^2 - M^2 \left (
4 {{m_f^2} \over {M^2}} - 2 {{(\gamma \cdot q)} \over M} {{m_f} \over M}
\right )
{{z/2v} \over {(1-z/2v)^2}} ~,
\label{fpot}
\eeq
where $m_f=h v$ is the physical mass of the fermion. The spin operator
$(\gamma \cdot q)/M$ in the rest frame of produced bosons is simply
$\gamma_0$. Therefore its eigenvalues in this frame are either $+1$ or $-1$
and a free fermion wave function at a non-zero spatial momentum contains
both eigenstates (upper and lower components of the Dirac bispinor in the
standard representation).  Therefore comparing with the previously discussed
bosonic case with spontaneously broken symmetry, we conclude that the
nullification of the threshold amplitudes for the scattering $f\, {\bar f}
\to n \, H$ for $n > N$ occurs when the coefficient takes exceptional values
for both eigenstates, i.e.  the following two relations hold simultaneously:

\beq
4{{m_f^2} \over {M^2}} + 2 {{m_f} \over M}=N(N+1)~~~~{\rm and}~~~~
4{{m_f^2} \over {M^2}} - 2 {{m_f} \over M}=N_1(N_1+1)
\eeq
with both $N$ and $N_1$ being integer. This obviously is equivalent to the
``magic" relation (\ref{fmagic}).

It is interesting to notice that unlike the case of Higgs boson production
by transversal vector bosons the result for fermions that the threshold
amplitudes are zero for $n > N$ still leaves potentially non-zero one
kinematically possible amplitude, i.e. for $n=N$, which corresponds to the
degenerate situation when the fermion and antifermion being exactly at rest
produce $N$ Higgs bosons all being also at rest. This amplitude however is
ruled out by the gamma-matrix structure, provided that the Yukawa
coupling is scalar as in eq.(\ref{finter}). Indeed in this case in the
absence of any spatial momenta the gamma-matrix structure of the
amplitude of the scattering $2 \to N$ in the rest frame can only be given by
$A\,\gamma_0 +B$ with $A$ and $B$ being ordinary numbers. On the other hand
the bispinors for the static fermion and antifermion are orthogonal
different eigenvectors of the matrix $\gamma_0$ (the fermion bispinor has
only upper components non-zero, while in that for the antifermion non-zero
are the lower components). Therefore the amplitude is vanishing for the
on-mass-shell fermion-antifermion pair. \\[0.2in]

It is not known at present to what extent these unexpected infinite
series of zeros of the production amplitudes at the thresholds are affected
by the loop corrections. Neither it is clear whether this unexpected
behavior is a manifestation of a hidden symmetry.

I am thankful to L.B. Okun and V.A. Novikov for bringing to my attention
the papers \cite{zero2} where the threshold nullification to the leading
order in $m_V/m_H$ of the amplitudes of the scattering $V_L\,V_L \to H\,H$
was found by a conventional calculation of Feynman graphs. This work is
supported in part by the DOE grant DE-AC02-83ER40105.

{\Large \bf Figure captions}\\[0.3in]
{\bf Fig. 1}. The propagator $D_n(p)$ with emission of $n$ on-mass-shell
particles all being at rest. The circle represents the sum of all tree
graphs. \\[0.15in]
{\bf Fig. 2}. The recursion equation (\ref{recur}) for the propagators
$D_n(p)$. The filled circle corresponds to the sum of all tree graphs
originating from one-fold interaction, described by the vertex function
$V(\phi)$, of the propagating particle with the field $\phi$.

\newpage
\unitlength=1.00mm
\thicklines
\begin{picture}(100.00,131.00)(0.00,20.00)
\put(75.00,114.00){\circle{14.00}}
\put(50.00,114.00){\vector(1,0){18.00}}
\put(82.00,114.00){\vector(1,0){18.00}}
\put(75.00,121.00){\line(0,1){6.00}}
\put(78.00,120.00){\line(1,2){3.00}}
\put(72.00,120.00){\line(-1,2){3.00}}
\put(75.00,131.00){\makebox(0,0)[cc]{{\it \large $n$}}}
\put(90.00,110.00){\makebox(0,0)[cc]{{\it \large $p$}}}
\put(58.00,110.00){\makebox(0,0)[cc]{{\it \large $p + nq$}}}
\put(75.00,90.00){\makebox(0,0)[cc]{{\bf Figure 1}}}
\end{picture}

\begin{picture}(130.00,66.00)(0.00,-20.00)
\put(48.00,49.00){\circle{14.00}}
\put(48.00,56.00){\line(0,1){6.00}}
\put(51.00,55.00){\line(1,2){3.00}}
\put(45.00,55.00){\line(-1,2){3.00}}
\put(48.00,66.00){\makebox(0,0)[cc]{{\it \large $n$}}}
\put(63.00,45.00){\makebox(0,0)[cc]{{\it \large $p$}}}
\put(113.00,49.00){\circle{14.00}}
\put(113.00,56.00){\line(0,1){6.00}}
\put(116.00,55.00){\line(1,2){3.00}}
\put(110.00,55.00){\line(-1,2){3.00}}
\put(113.00,66.00){\makebox(0,0)[cc]{{\it \large $n-n_1$}}}
\put(128.00,45.00){\makebox(0,0)[cc]{{\it \large $p$}}}
\put(55.00,49.00){\vector(1,0){11.00}}
\put(120.00,49.00){\vector(1,0){10.00}}
\put(28.00,49.00){\vector(1,0){13.00}}
\put(78.00,50.00){\makebox(0,0)[cc]{{\LARGE $= \Sigma$}}}
\put(75.00,15.00){\makebox(0,0)[cc]{{\bf Figure 2}}}
\put(86.00,49.00){\vector(1,0){20.00}}
\put(95.00,49.00){\circle*{5.20}}
\put(95.00,51.00){\line(0,1){6.00}}
\put(96.00,51.00){\line(3,5){3.00}}
\put(94.00,51.00){\line(-3,5){3.00}}
\put(95.00,61.00){\makebox(0,0)[cc]{{\large \bf $n_1$}}}
\put(95.00,42.00){\makebox(0,0)[cc]{{\large \bf $V$}}}
\put(83.00,45.00){\makebox(0,0)[cc]{{$n_1$}}}
\end{picture}

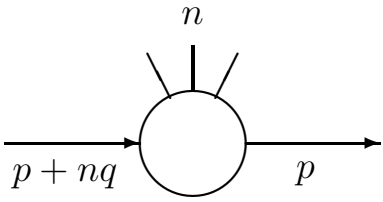
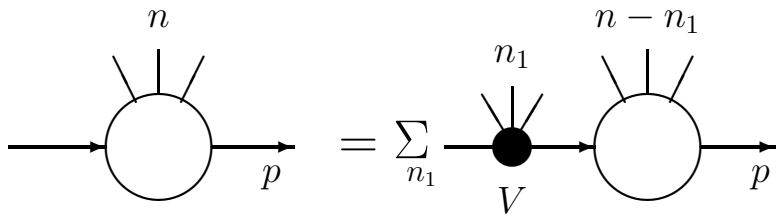
\end{document}